\documentclass[twoside,11pt]{article}

\usepackage{jmlr2e}

\usepackage[dvipsnames]{xcolor}
\definecolor{codebg}{rgb}{0.98,0.92,0.92} 

\usepackage[frozencache,cachedir=mintedcache]{minted}

\setminted{
    fontsize=\small,
    linenos,             
    frame=lines,         
    framesep=2mm         
}

\newcommand{\ModelName}{\raisebox{-0.5ex}{\includegraphics[height=2.7ex]{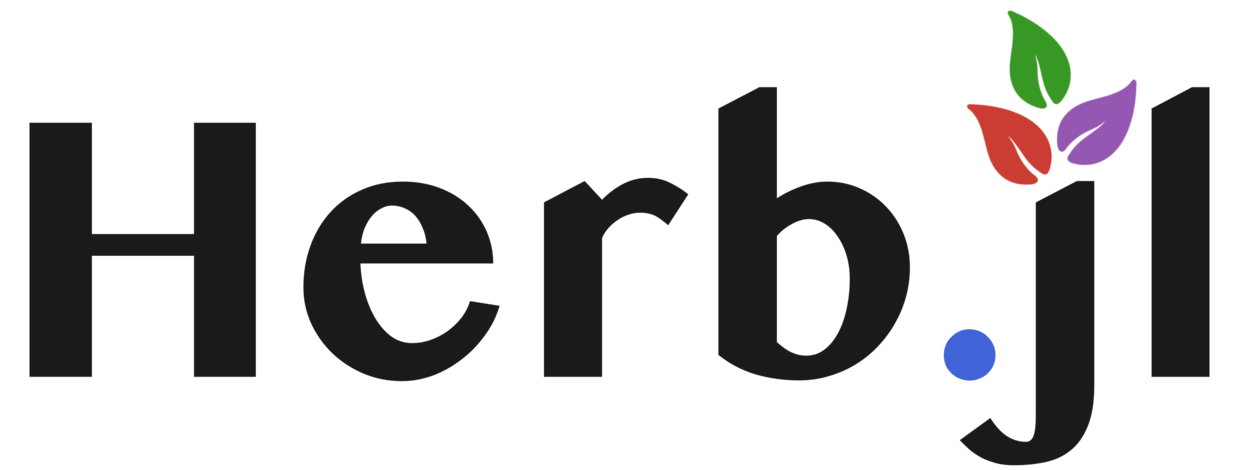}}}

\usepackage{todonotes}

\usepackage{subcaption}
\usetikzlibrary{fit}
\usepackage{amsmath}

\usepackage{forest}
\forestset{
  dashedbox/.style={draw,dashed,inner sep=2pt}
}

\usepackage{lastpage}
\jmlrheading{01}{2026}{1-\pageref{LastPage}}{01/26}{--/--}{Hinnerichs et al.}

\ShortHeadings{Herb.jl: A unifying Program Synthesis Library}{Hinnerichs et al.}
\firstpageno{1}

\begin{document}

\title{Extended Paper: An Introduction to Herb.jl: A Unifying Program Synthesis Library}

\author{\name Tilman Hinnerichs \email t.r.hinnerichs@tudelft.nl
       \AND
       \name Reuben Gardos Reid \email r.j.gardosreid@tudelft.nl
       \AND
       \name Jaap de Jong \email J.dejong-18@student.tudelft.nl
       \AND
       \name Bart Swinkels \email B.J.A.Swinkels@student.tudelft.nl
       \AND
       \name Pamela Wochner \email p.wochner@tudelft.nl
       \AND
       \name Issa Hanou \email i.k.hanou@tudelft.nl
       \AND
       \name Nicolae Filat \email n.filat@student.tudelft.nl
       \AND
       \name Tudor Magurescu \email magirescu@student.tudelft.nl
       \AND
       \name Sebastijan Dumancic \email s.dumancic@tudelft.nl \\
       \addr Technical University Delft, Delft, Netherlands}

\editor{--}

\maketitle

\begin{abstract}
Program synthesis -- the automatic generation of code given a specification -- is one of the most fundamental tasks in artificial intelligence (AI) and the dream of many programmers. 
Numerous synthesizers have been developed to tackle program synthesis, offering different approaches to the exponentially growing program space.
Although numerous state-of-the-art program synthesis tools exist, reusing and adapting them remains tedious and time-consuming. 
We propose Herb.jl, a unifying program synthesis library written in Julia, to address these issues. 
Since current methods share similar building blocks, we aim to break down the underlying algorithms into extendable, reusable subcomponents. 
To demonstrate the benefits of using Herb.jl, we show three common use cases:
1.\@ How to implement a simple problem and grammar, and how to solve it, 
2.\@ How to implement and customize a previously developed synthesizer with just a few lines of code, 
and 3.\@ How to run a synthesizer against a benchmark.
\end{abstract}

\begin{keywords}
  Program Synthesis, Constraint Programming, Reproducible Research, Neural-Symbolic AI
\end{keywords}

\section{Introduction}
Many programmers dream that a computer automatically writes its own code at their bidding: ideally, simply stating ``fix this failing test'' or ``write this function for me'' would save developers many hours and headaches. 
Given the \textit{specification}, a precise description of what the program should do, and a \textit{program space}, essentially a description of the target programming language, an algorithm could then automatically generate the desired implementation. 
This ability to go from specification to code is the goal of program synthesis.

Many problems across computer science and beyond can be formulated as program synthesis problems~\citep{kuniyoshi1994learning,kuncak2012software,Solar-Lezama13,SemGuS,SyGuS2019,ChaudhuriNeurosymbolicProgramming}.
As an example, test-driven development can be viewed as a synthesis task: the tests serve as the specification, and the implementation language chosen by the client defines the program space. 
Synthesizers can thus be used to write provably correct programs from scratch, or fix or optimize existing code snippets~\citep{PhothilimthanaT16Superoptimization,Monperrus18ProgramRepair}.
Program synthesis has found applications beyond software engineering, in discovering new algorithms~\citep{DimensionsinPS}, in education and teaching~\citep{HeadGSSFDH17}, modelling complex domains in biology and chemistry~\citep{Higuera05CompBio,SantosNPMS12Biology,GulwaniPS17}, and human--computer interaction~\citep{ButlerTP17}.

Program synthesis is a hard problem.
Often framed as a search problem, the synthesizer searches over all possible programs to find the one that satisfies the specification.
Unfortunately, the space of programs is infinite and grows exponentially with the program depth, rendering naive search infeasible.
Finding ways to restrict and efficiently traverse the program space is crucial to tackling real-world problems with program synthesis.

Numerous program synthesis approaches have been developed~\citep{GulwaniPS17,ChaudhuriNeurosymbolicProgramming}.
Many synthesizers follow common principles to simplify the problem: for example, they either restrict the program space by limiting program complexity or by imposing constraints~\citep{Neo,hinnerichs2025modelling}, solve sub-problems and recombine their solutions~\citep{EUSolver}, or use heuristics to guide the search~\citep{Balog2017DeepCoder,Probe2020Barke}.

However, developing new synthesizers that build upon state-of-the-art implementations is tedious.
The existing synthesizer implementations are often tightly coupled to publications, resulting in the code engineered to run only the specific method on a specific set of benchmarks.
This lack of planning for reuse results in the constant need to reimplement the basic functionality shared by most program synthesizers. 
This lack of modularity and reuse makes it difficult to compare ideas or transfer techniques across different implementations.

We identify four recurring problems that often arise when reusing or adapting implementations of existing synthesizers by, for example, changing a small detail like a heuristic or applying it to a new domain.

\begin{enumerate}
    \item \textbf{Synthesizer implementations are domain-specific and hard to adapt to other domains.}
    \item \textbf{Synthesizers consist of the same building blocks that cannot be reused easily in practice}.
    \item \textbf{Synthesizers are hard to compare due to varying engineering choices.}
    \item \textbf{Synthesizer's benchmarks are hard to reuse and rerun.}
\end{enumerate}
We provide a detailed description of each problem, along with a motivating example, in Section~\ref{sec:Motivation}.
We argue that each problem hinders progression and unification in the field of program synthesis.

\vspace{0.25cm}

To address these issues, we propose \ModelName{}, a general-purpose program synthesis library. 
\ModelName{} aims to support both novices in using program synthesisers and the experts developing new techniques. 
We identify the common building blocks shared across the wide range of existing program synthesizers and provide a modular, extensible, and reusable implementation, allowing straightforward reapplication or recombination of existing methods. 
More precisely, in \ModelName{}, one can formulate a program synthesis problem by providing a specification to describe the desired functionality of a target program, a grammar to define a program space, a program interpreter capable of running the program defined by the grammar, constraints on program structure, and a search procedure to traverse the program space.
While we provide default implementations for each of these components, they can be modified or replaced with custom implementations. 
Finally, we provide a unified re-implementation of standard benchmarks in a human-readable and adaptable form in \texttt{HerbBenchmarks.jl}.


Using these modular building blocks, we have implemented a range of well-known synthesizers in \texttt{Garden.jl}.
\texttt{Garden.jl} aims to be a repository of reference implementations that support the program synthesis community in fairer and easier benchmarking.
Moreover, these implementations are not only generalised to a variety of program synthesis formulations but are also easily customizable. 

In summary, this paper contributes:  
\begin{itemize}
    \item \ModelName{}, a unifying program synthesis library  in the Julia programming language,
    \item a series of demonstrations on how to easily implement previously developed synthesizers using \ModelName{},
    \item a range of standard benchmarks in human-readable and extendable format, and
    \item an overview of guiding design principles in \ModelName{}, relevant to implementations of other synthesizers within and outside of \ModelName{}.
\end{itemize}

The rest of this paper is structured as follows:
We introduce the problem of program synthesis and common terminology in Section~\ref{sec:background}.
Second, we identify four common problems that arise when trying to reuse and build upon existing program synthesis works in Section~\ref{sec:Motivation}.
We then present \ModelName{} in two ways:
Section~\ref{sec:using_Herb} is intended for users who want to apply \ModelName{} by formulating and solving their own program synthesis problems.
Section~\ref{sec:developing_Herb} is intended for developers who want to build upon, customize, and extend \ModelName{}'s functionality.



\section{Program Synthesis}
\label{sec:background}
We briefly introduce program synthesis and common terminology.

A program synthesis problem is defined by two components: (1) a \emph{specification}, describing the user’s intent, and (2) a \emph{grammar}, describing the target language.  
A common way to express a specification is through input–output (IO) examples, which the synthesized program must satisfy.  
The grammar is composed of a set of context-free derivation rules that define the syntax, i.e., the structure of all valid programs in the target language.

\begin{example}[Getting Started]
\label{ex:integer_arith}
As a running example, consider a simple integer arithmetic domain.  
The grammar contains integer literals~\texttt{1, 2, ...}, binary operations~\texttt{+} and~\texttt{*}, and a variable symbol~\texttt{x} representing the input:  

\[
\begin{array}{ll}
\text{Int} = & \texttt{x}\\
\text{Int} = & \texttt{1}\\
\text{Int} = & \texttt{2}\\
\text{Int} = & \texttt{...}\\
\text{Int} = & \text{Int}~\texttt{+}~\text{Int}\\
\text{Int} = & \text{Int}~\texttt{*}~\text{Int}\\
\end{array}
\]

The specification is given by a set of input–output examples:
\[
\mathcal{E} = \{(0,1), (1,3), (2,5), (3,7)\}.
\]
\end{example}


We represent programs as abstract syntax trees (ASTs), where nodes are either terminal symbols (i.e., concrete functions and values) or non-terminal symbols (i.e., parts of the program that still need to be filled in).
Leaves corresponding to non-terminal symbols are referred to as \emph{holes}.  
Any program tree that still contains one or more holes is called a \emph{partial program}.  
When all holes have been replaced by terminal nodes, the program is said to be \emph{complete}.

\begin{example}[Partial Program]
The program
\[
\texttt{ (x + Int) + Int}
\]
contains two holes with non-terminals $\{\texttt{Int}, \texttt{Int}\}$, that still need to be filled, and is thus partial.
\end{example}

Various search strategies exist to efficiently traverse the space of possible programs. 
In general, search methods can be categorized into three families:
A \emph{top-down search} begins with the start symbol of the grammar as the root node, and iteratively tries to fill the holes by applying derivation rules from the grammar.
A \emph{bottom-up search} starts from complete programs and combines concrete programs to construct bigger ones according to derivation rules. 
A \emph{stochastic search}, like genetic programming and Markov-Chain-Monte-Carlo methods (MCMC), samples an initial program and repetitively modifies the program by removing and re-sampling parts of the program.

\section{Motivation}
\label{sec:Motivation}

We outline common problems that the community faces when reusing existing implementations.

\subsection{A Motivating Example}
\label{sec:Motivation:Example}

We introduce a motivating example to illustrate these problems:

\begin{example}[A New Approach]
    \label{ex:introduction}
    Assume we want to implement software for low-level hardware, such as firmware for a hardware controller, which is notoriously difficult to write.
    We have chosen a small, performance-oriented programming language that closely maps to the microcontroller’s instruction set, making it ideal for this task but tedious and error-prone to program by hand. 
    Instead, we would like to use a synthesis tool to generate code in this language from a given set of tests (our specification), i.e., we aim to use program synthesis on \textbf{a novel program synthesis domain}.

    Assume we also have a \textbf{novel idea for a heuristic}, for example, based on an estimate of the execution time of the program.
    This helps the synthesizer to prioritize useful programs that also comply with the specification.
\end{example}

In program synthesis, we use the term \emph{specification} to refer to what we want our resulting program to do, such as input-output examples. 
The language our program can use, often tailored to the specification, is described by the \emph{grammar}, which we will use interchangeably.
A \emph{heuristic} can be any function that takes a possible program and returns a score indicating the quality of the considered program. 
The scores are then used to guide the search, for example, by ranking solutions or pruning unwanted candidates.

\subsection{Why Existing Synthesizers Are Not Enough}
\label{sec:Motivation:Problems}

We outline the problems with reusing current program synthesis implementations.
Suppose we now try to use an existing synthesizer to solve a new synthesis problem described in Example~\ref{ex:introduction}.
Then we likely encounter one of the following problems.

\paragraph{Problem 1: Synthesizer implementations are domain-specific.}
While program synthesis works introduce general, widely applicable ideas, their implementations are often tailored to a particular domain and the experiments considered in the paper.
For instance, the grammar might be hard-coded, or the implementation might assume that the language comes with an SMT-LIB formalisation.
Formulating problems within the existing implementation can be challenging, especially for beginners, unless the two requirements align perfectly.

\paragraph{Problem 2: Synthesizers comprise the same building blocks that are not easily reusable in practice.}
To implement our heuristic idea, as presented in Example~\ref{ex:introduction}, we would ideally be able to utilize the building blocks from previous implementations, such as a cost function, and plug them into our new implementation. 
Unfortunately, existing implementations are rarely modular, and they often slightly alter the formalization of the synthesis problem.
With synthesizers tailored to a specific approach, researchers have to re-implement the same ideas repeatedly just to compare them.
Furthermore, the lack of modularity prevents cross-fertilisation of ideas from different branches of program synthesis.
For example, extending a constraint-based synthesizer with a simple heuristic can be challenging.



\paragraph{Problem 3: Synthesizers are hard to compare due to varying engineering choices.}
Many approaches do not make their assumptions and optimizations explicit, such as the choice of data structures, cached program evaluation, parallelization, or simply using a faster programming language. 
As a result, empirical comparisons can easily conflate algorithmic improvements with engineering effort.
Suppose we implement a novel heuristic idea within a new synthesizer and compare its performance with that of another heuristic implemented in a different programming language.
Even if one tool consistently outperforms the other, it is difficult to determine whether this reflects a superior heuristic or a more effective implementation. 
For instance, a tool written in a lower-level programming language with carefully engineered caching may achieve significantly lower runtimes, making a weaker heuristic appear competitive with, or even superior to, a stronger one.

\paragraph{Problem 4: Benchmarks are hard to use.}
Assume we have our solution to Example~\ref{ex:introduction}, which seems to be an effective synthesizer. 
To compare this approach with others, we also need a benchmark that ensures a fair comparison: a set of program synthesis problems defined by a specification and a corresponding grammar.
While most works keep the specification the same, they often modify the grammar.
This change, however, has important consequences for the evaluation, as such grammar changes alter the program space, which dictates the difficulty of a synthesis problem.
Hence, while many works run the same benchmark, they solve vastly different problems.

\section{Using and Applying Herb.jl}
\label{sec:using_Herb}
This section is intended for users who want to apply \ModelName{} by formulating their own program synthesis problems.

\begin{figure}[h]
    \centering
    \includegraphics[width=0.8\textwidth]{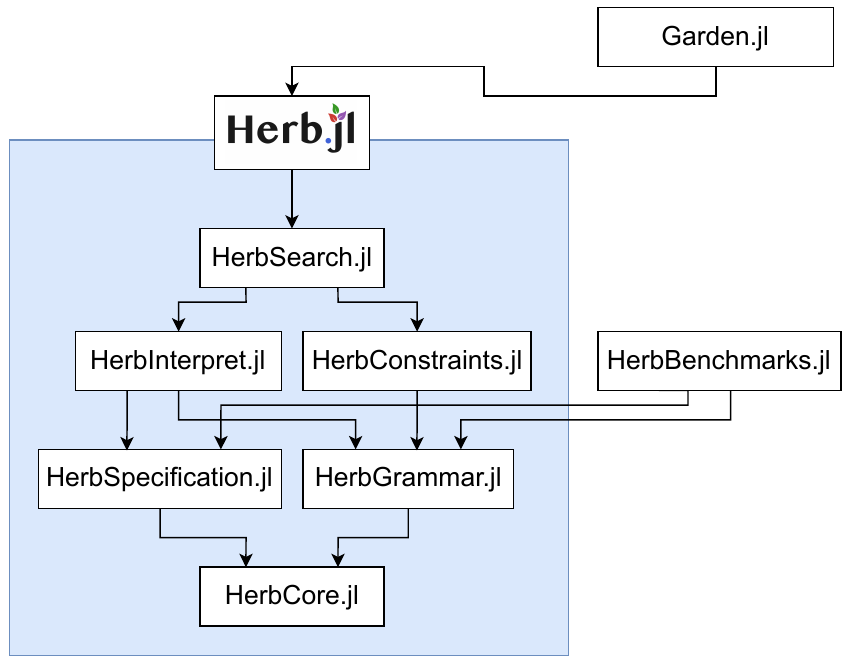}
    \caption{
        \textbf{Dependency graph of packages in the \ModelName{} ecosystem.}
        Both \texttt{HerbBenchmarks.jl} and \texttt{Garden.jl} are separate packages. 
        } 
    \label{fig:dep_graph}
\end{figure}

Reusing current synthesisers is often not easy in practice, as highlighted by the problems in Section~\ref{sec:Motivation}.
We focus on global ease of use and modularity across the entire \ModelName{} framework.
To facilitate modularity, \ModelName{} comprises multiple submodules that enable straightforward composition of common functionality. 
We first outline the core components (shown in Figure~\ref{fig:dep_graph}) with an exemplary program synthesis workflow.
Second, we introduce two special modules implemented using \ModelName{} that showcase interoperability: \texttt{Garden.jl}, a collection of synthesisers, and \texttt{HerbBenchmarks.jl}, a collection of program synthesis benchmarks and problems.

\paragraph{Software availability}
All listed component packages are available from the Julia package repository. \footnote{\url{https://juliapackages.com/p/herb}}
\ModelName{} acts as an umbrella package that brings all components together and is the likely entry point for most users.
For development, all packages are also available as individual GitHub repositories.\footnote{\url{https://github.com/Herb-AI/Herb.jl}}
Note that \texttt{Garden.jl} and \texttt{HerbBenchmarks.jl} are separate repositories and need to be loaded explicitly.

\subsection{An Exemplary Program Synthesis Workflow}
\label{sec:PS_workflow}

We demonstrate how to express Example~\ref{ex:integer_arith} in \ModelName{} and search the program space with a simple search method. 
This requires defining
\begin{itemize}
    \item the problem specification,
    \item the program space, and
    \item a search method.
\end{itemize}

\paragraph{Defining the core.}
The core definitions and interfaces are defined in \texttt{HerbCore.jl} module. 
This module defines interfaces for grammars, programs, and constraints and how to interact with them, providing the abstract types and functions that the rest of the \texttt{Herb*} packages build upon.
For example, it defines the struct \texttt{RuleNode}, which represents an AST (i.e., a program).

\paragraph{Defining a problem.}
We express the problem specification (see Example~\ref{ex:integer_arith}) as a set of input--output examples.
Each example pairs an input dictionary that maps variables to values with the desired output.
\texttt{HerbSpecification.jl} provides functionality for defining a range of specification types: 

\begin{minted}{julia}
using HerbSpecification
problem = Problem([
	IOExample(Dict(:x => 0), 1),   # :x is the input symbol
	IOExample(Dict(:x => 1), 3),
	IOExample(Dict(:x => 2), 5),
	IOExample(Dict(:x => 3), 7) 
])
\end{minted}

\paragraph{Defining the program space} 
In addition to defining the problem itself, we must also define the syntax of valid programs that constitute possible solutions. 
Conceptually, we use context-free grammars (CFGs) to specify the space of programs to be synthesized.

A grammar consists of a set of rules, each of which is either (1) a terminal rule, representing a constant or input variable of a given type, or (2) a derivation rule, which specifies how abstract syntax trees (ASTs) of certain types can be composed into larger expressions of a given type.

in \ModelName{}, grammars are written using \texttt{@csgrammar} macro from \texttt{HerbGrammar.jl}.
The following definition specifies a grammar for integer arithmetic:

\begin{minted}{julia}
using HerbGrammar
grammar = @csgrammar begin
    Int = 1 | 2 | x   # Rule indices 1, 2, 3
    Int = Int + Int   # Rule 4
    Int = Int * Int   # Rule 5
end 
\end{minted}
Here, $x$ denotes the input symbol, which defines the variable $x$. $x$, along with the constants $1$ and $2$ are the terminal rules in this grammar. 
\texttt{Int = 1|2|x} is a short-hand for writing 3 rules in one line. Then, there are two derivation rules, defining how two \texttt{Int}s can be added or multiplied to define a new \texttt{Int}, respectively. 

While this grammar only contains one type, \texttt{Int}, and one input, we can generally define multiple types and inputs.
To add rules with a new type \texttt{Bool} to the existing \texttt{grammar}, we can use, for example an if-then-else rule

\begin{minted}{julia}
add_rule!(grammar, :(Int = Bool ? Int : Int))
add_rule!(grammar, :(Bool = Int == Int))
\end{minted}

A core modelling tool for program spaces in \ModelName{} is constraints, which constrain the syntax of the programs explored.
Constraints, implemented in \texttt{HerbConstraints.jl},  allows us to refine program spaces in a way that is not possible with context-free grammars, which permit many syntactically valid but semantically redundant or irrelevant programs.
For example, we can break the commutative symmetry of the addition operator $+$ using the \texttt{Ordered} constraint, which states that the left operand of $+$ needs to be a program that is lexicographically smaller than the right operand, and forbid the combination $1*Int$ using the \texttt{Forbidden} constraint, as follows:

\begin{minted}{julia}
using HerbConstraints

# Remove symmetry over Int + Int
template_tree = RuleNode(4, [VarNode(:a), VarNode(:b)])
order = [:a, :b]
ord_con = Ordered(template_tree, order)
addconstraint!(grammar, ord_con)

# Remove 1 * Int 
# Index 5 is *, index 1 is 1
forbid_con = Forbidden(RuleNode(5, [RuleNode(1), VarNode(:A)])) 
addconstraint!(grammar, fordbid_con)
\end{minted}
Here, $4$ denotes the index of the addition rule within the grammar.
For a full description of our constraint system and the underlying constraint solver, please see \citet{hinnerichs2025modelling}.


\paragraph{Using an existing search technique}
Lastly, we have to define how to traverse the program space using a search technique.
To search for the solution program, \texttt{HerbSearch.jl} defines standard search procedures, in the form of \textit{iterators}, and a toolbox for creating new search techniques 

For example, we can use off-the-shelf iterators, like breadth-first search (BFS), to enumerate programs by increasing program size: 
\begin{minted}{julia}
using HerbSearch
iterator = BFSIterator(grammar, :Int, max_depth=5)
solution, flag = synth(problem, iterator) # flag is optimal_program
println(solution) # yields 4{3,4{1,3}}
\end{minted}

The search requires the starting symbol \texttt{:Int}, which defines the root non-terminal for deriving candidate programs. 
We can also provide an optional stopping criterion, e.g., one that limits iterations to programs with a maximum depth of 5. 
The \texttt{synth} function executes the search method on the problem, returning the solution and a status flag indicating whether the search was successful. 
The \texttt{:optimal\_program} flag indicates the solution satisfies all examples.

\ModelName{} represents programs as ASTs.
We can translate the solution to a Julia expression and evaluate it on a concrete input value using the module \texttt{HerbInterpret.jl}:
\begin{minted}{julia}
program = rulenode2expr(solution, grammar) # yields :(x + (1 + x))
output = execute_on_input(grammar, program, Dict(:x => 5)) # yields 11
\end{minted}

\paragraph{Conclusion}
This concludes our exemplary program synthesis workflow.
This use case demonstrates that \ModelName{} can express program synthesis problems and the search to find a solution in just a few lines of readable code. 
Note that any of the components, i.e., problem specification, grammar and search method, can be easily substituted and extended, which we discuss in Section~\ref{sec:developing_Herb}.

\subsection{Using an Existing Synthesizer}
\label{sec:garden}

\texttt{HerbSearch} provides the necessary functionality to implement synthesizers in a quick and modular fashion.
Using those building blocks, \ModelName{} provides concrete implementations of existing synthesizers from different synthesizer families, such as Probe~\citep{Probe2020Barke}, FrAngel~\citep{Shi2019frangel}, and Neo~\citep{Neo}.
These implementations are available from \texttt{Garden.jl},\footnote{\url{https://github.com/Herb-AI/Garden.jl}} a separate repository with reference implementations of existing synthesisers.
Moreover, \texttt{Garden.jl} also provides utilities to facilitate experimentation and comparison across synthesizers.
This separation keeps \texttt{HerbSearch} lightweight and focused on core abstractions, while allowing \texttt{Garden.jl} to evolve as a community-driven collection of synthesizers.

For example, to run Probe on a problem, we can use
\begin{minted}{julia}
using Garden
solution = probe(grammar, :Int, problem; probe_cycles = 3, max_depth=5) 
\end{minted}
which runs Probe for 3 cycles using a cost-based bottom-up search constrained to programs of maximum depth of 5.


\subsection{Using a Benchmark}
\label{sec:benchmark}
\ModelName{} collects various program synthesis benchmarks in \texttt{HerbBenchmarks.jl} and standardises them in a common format.
Currently, the module includes the SyGuS challenge~\citep{SyGuS2019}, the ARC challenge~\citep{chollet2019arc}, the Karel dataset~\citep{pattis1994karel}, and problem sets from  including common program synthesis benchmarks from DeepCoder~\citep{Balog2017DeepCoder,Neo} and BRUTE~\citep{Brute}. 

All benchmarks are written in \ModelName{}'s syntax and are easy to use out of the box.
For example, to load the string transformation tasks from the SyGuS challenge 2019~\citep{SyGuS2019}, we can use
\begin{minted}{julia}
using HerbBenchmarks
pairs = get_all_problem_grammar_pairs(PBE_SLIA_Track_2019) 
\end{minted}

To evaluate an existing synthesizer, e.g., Probe, on all problems, we can iterate over all problem-grammar pairs:

\begin{minted}{julia}
solved_problems = 0
for (problem, grammar) in pairs
    @time solution = probe(grammar, :Start, problem; max_depth=7)
    if !isnothing(solution)
        solved_problems += 1
    end
end
\end{minted}
This snippet counts the number of solved problems. 
Here, we use the Julia macro \texttt{@time} to output the computing resources, such as execution time and memory usage, for each problem. 
Additional measures can be easily included using standard Julia libraries like BenchmarkTools.jl\footnote{\url{https://github.com/JuliaCI/BenchmarkTools.jl}}.

\section{Developing and Customizing Herb.jl}
\label{sec:developing_Herb}
This section is intended for developers who want to build upon, customise and extend \ModelName{}'s functionality.

We give an overview of common interfaces in \ModelName{}.
We first describe how to customise search methods and define custom program search orders over programs.
Second, we describe how to implement an existing synthesizer, namely Probe, with a few lines of code.

We provide more examples and guides for customisation in our documentation~\footnote{https://herb-ai.github.io/Herb.jl/dev/}.

\subsection{Defining custom program space traversal}

Guided search is one of the most common strategies for navigating the vast space of programs~\citep{GulwaniPS17,HeapSearch,Brute,CrossBeam,AbstractBeam,BeeSearch,EcoSearch}.
Here, the user defines a heuristic that prioritises more promising programs.

\paragraph{Syntax and semantics}
\ModelName{} follows the syntax-guided synthesis~\cite{SyGuSAlur} paradigm, in which the syntax and semantics are provided separately. 
The search procedures operate purely syntactically, by changing an abstract syntax tree of explore new programs. 
The user also provides a separate program interpreter to evaluate the constructed programs.
However, the search procedures in \ModelName{} only care about the syntax.


\paragraph{Uniform Trees}
\label{sec:uniform_trees}

\ModelName{} uses a custom data structure to represent 
programs, called \textit{uniform trees}.
Uniform trees represent multiple programs of similar \textit{syntactic structure}, i.e., programs whose ASTs have the same \textit{shape} -- the ASTs have the same number of nodes, the same connectivity between the nodes, and nodes of the matching type (see Figure~\ref{fig:uniform} for an example of a uniform tree, and  \citet{hinnerichs2025modelling} for a full description).
While this design choice is almost entirely hidden from the casual user, it has an important impact on developers:
uniform trees are stateful data structures in which the values at the nodes can change.
The program space iterators will reuse the same data object when exploring different programs, as long as they fit the same uniform tree.
Therefore, if a developer wants to store an explored program for future reference, they must explicitly copy the desired AST from the uniform tree, as the reference data object will change.

\paragraph{Constraint solver}
The second important component in \ModelName{} is the constraint solver, which ensures that the explored program satisfies the user-imposed syntactic constraints.
These constraints are used to eliminate redundant programs (i.e., multiplying any number by 1) or those that one knows cannot be a part of the solution (i.e., a program that does not use any of the provided arguments). 
Constraint solver is, in fact, in charge of changing the state of uniform trees; the program space iterators delegate those changes to it.
This solver is responsible for tracking and propagating constraints, ensuring that violating programs are never encountered during the search. 

Most importantly, \ModelName{} already implements this constraint handling for the three classes of synthesizers: top-down, bottom-up, and stochastic search.

\begin{figure}[h]
    \centering
    \begin{forest}
        for tree={
          s sep=7mm,
          l sep=4mm,
          align=center
        }
        [$\times$
          [{$\{+,\times\}$}, dashedbox
            [$1$]
            [{$\{1,x\}$}, dashedbox]
          ]
          [$+$
            [$1$]
            [$1$]
          ]
        ]
    \end{forest} 
    \caption{Example of a uniform tree, where the dashed boxes indicate uniform nodes, i.e., operators that have the same `shape': They have the same number of children with the same type. 
    }
    \label{fig:uniform}
\end{figure}
 

\paragraph{Defining program search order}
Most guided search algorithms rely on a default data structure that determines the next step in the search. 
For example, top-down and bottom-up search use a priority queue, whereas stochastic and genetic algorithms determine the next program by randomly mutating the current one.

To customise search order in \ModelName{}, we have to define which \textit{uniform tree} to search next, and which program \textit{within} a uniform tree to enumerate next.

In top-down search, the main interface for developers is the priority queue, where each uniform tree's priority is defined by a \texttt{priority\_function}.
Within a uniform tree, the order in which to enumerate the elements of a node's domain is defined using a \texttt{derivation\_heuristic}.
Overwriting the priority function yields different search techniques: decreasing the priority of new elements implements a depth-first search (DFS), while increasing the priority leads to breadth-first search (BFS) behaviour. 

Our bottom-up search implementation provides three main interfaces for customisation.
First, bottom-up searches maintain a bank of seen programs, indexed by a \textit{measure} (e.g., size, depth, or cost), allowing for any numeric type.
New programs are formed by combining simpler ones already in the bank; the user can customise the \texttt{combine} interface to control how new programs are constructed. 
Second, after iterating a program, we can decide whether to keep it and store it in the bank using \texttt{add\_to\_bank}. 
Third, the priority values in the priority queue and the program's measure are calculated using \texttt{calc\_measure}, which takes a program as input and returns its measure.

We concretely implement a cost-based search for both top-down and bottom-up search.
For top-down search, given a probabilistic grammar, we can implement a most-likely-first search (\texttt{MLFSIterator}) that enumerates programs with the highest probability first.
We first define a new type, and then customise the two priority functions.
\begin{minted}{julia}
@programiterator MLFSIterator() <: TopDownIterator

function priority_function(
    ::MLFSIterator,
    grammar::AbstractGrammar,
    current_program::AbstractRuleNode,
    parent_value::Union{Real,Tuple{Vararg{Real}}}
)
    -max_rulenode_log_probability(current_program, grammar)
end

function derivation_heuristic(iter::MLFSIterator, domain::Vector{Int})
    log_probs = get_grammar(iter).log_probabilities
    return sort(domain, by=i -> log_probs[i], rev=true) # highest probability first
end 
\end{minted}

Similarly, we can redefine the search order for bottom-up search methods to implement a cost-based bottom-up search.
\ModelName{}’s bank abstraction is generic over the measure type, so integer measures (e.g., AST size) and floating-point measures (e.g., cost) can be easily swapped.
We first define the new iterator type using a custom measure type. 
Since costs, e.g. in Probe, are floating-point values derived from log-probabilities, we instantiate the bank accordingly:
\begin{minted}{julia}
@programiterator CostBasedBottomUpIterator(
    bank=MeasureHashedBank{Float64}(),     # maps cost to programs
    max_cost::Float64=Inf,                 # global cost cap
    current_costs::Vector{Float64}=Float64[],
) <: AbstractCostBasedBottomUpIterator 
\end{minted}

Next, we overwrite the default measure calculation to use negative log-probability, prioritising programs with lower costs, similar to \texttt{MLFSIterator}.
\begin{minted}{julia}
function calc_measure(iter::CostBasedBottomUpIterator, rn::AbstractRuleNode) 
    -rulenode_log_probability(rn, get_grammar(iter))
end
\end{minted}

\paragraph{Overwriting functions in Julia}

\ModelName{} is implemented in the Julia programming language for a range of reasons.
Julia relies on \textit{multiple dispatch}, where, simply put, the function definition can be dynamically determined at run time \citep{JuliaDispatch2018Bezanson}.
Multiple dispatch enables us to extend existing functions to new types (core to object-oriented programming) and allows us to add new functions to existing types (core to functional programming).
As a result, developers can easily add and overwrite new functions, like \texttt{calc\_measure}, and types, like \texttt{MLFSIterator}.


\subsection{(Re-)Implementing a Synthesizer}
We use \ModelName{} to re-implement Probe~\citep{Probe2020Barke}, showing how to write and customise basic functionality.

The Probe algorithm performs multiple so-called \textit{cycles}.
In each cycle, Probe (1) enumerates programs, (2) collects promising candidates with their fitness scores, and (3) updates the grammar rule probabilities. 
Given a problem, a (probabilistic) grammar, and a starting symbol, we can implement a cycle as follows.
\begin{minted}{julia}
function probe(
        grammar::AbstractGrammar,
        starting_sym::Symbol,
        problem::Problem;
        probe_cycles::Int = 3,
        kwargs...
    for _ in 1:probe_cycles
        # 1. Enumerate programs 
        iter = CostBasedBottomUpIterator(
            max_cost      = max_cost,
            current_costs = get_costs(grammar);  # -log p(rule)
            kwargs...
        )
        # 2. Collect promising programs
        # Returns (program, score) pairs and the result flag
        promising_programs, result_flag = get_promising_programs_with_fitness(
            iterator, problem)
        
        if result_flag == optimal_program
            program, score = only(promising_programs) # returns only element
            return program
        end

        # 3. Update rule probabilities
        modify_grammar_probe!(promising_programs, grammar)
    end
    return nothing # no solution found
end
\end{minted}

\paragraph{Customizing program evaluation}
\ModelName{} provides the standard \texttt{HerbSearch.synth} function to run iterators and return the best program. 
\texttt{synth} should be viewed only as a template; we encourage all users and developers of \ModelName{} to write custom evaluation functions, as custom synthesisers require custom information.

For Probe, we need to collect useful programs along with their corresponding fitness scores. 
Instead of \texttt{HerbSearch.synth}, we thus use the custom \texttt{get\_promising\_programs\_with\_scores} to achieve this. 
We enumerate over all programs in the iterator and evaluate them on the given problem.
If a program solves at least one example from our problem, we store it as a promising program.
\begin{minted}{julia}
function get_promising_programs_with_fitness(
        iterator::ProgramIterator,
        problem::Problem;
        max_time = typemax(Int),
        max_enumerations = typemax(Int),
        mod::Module = Main
)
    ...
    promising_programs = Set{Tuple{RuleNode, Real}}()

    for (i, candidate_program) in enumerate(iterator)
        # Decide whether to keep the program
        # Returns the share of examples solved. 
        fitness = decide_probe(candidate_program, problem, grammar, symboltable)

        if fitness == 1  # all examples solved
            push!(promising_programs, (freeze_state(candidate_program), fitness))
            return (promising_programs, optimal_program)
        elseif fitness > 0  # some examples solved
            push!(promising_programs, (freeze_state(candidate_program), fitness))
        end

        # Check stopping criteria
        if i > max_enumerations || time() - start_time > max_time
            break
        end
    end
    return (promising_programs, suboptimal_program)
end
\end{minted}
Here, \texttt{decide\_probe} returns the fraction of successfully solved examples.
The full code is available in the Garden (see Section \ref{sec:garden}).

\section{Conclusion}

\ModelName{} tackles a common problem in the field of program synthesis:
How can we make existing approaches reusable for a new use case and easily extend them with a novel idea? 
We outline common challenges researchers encounter when applying, extending, or comparing program synthesizers. 

We introduce \ModelName{}, a program synthesis library written in Julia that defines standard building blocks in program synthesis, allowing us to evaluate it against a standard set of benchmark problems. 
Moreover, \ModelName{} implements a range of off-the-shelf synthesizers.
\ModelName{} exploits Julia's meta-programming and dispatch features. 
This allows for a fast, stable, and adaptable framework.

We highlight three example use cases and how to tackle them using \ModelName{}.
We aim to motivate researchers in program synthesis to use \ModelName{} to make their ideas accessible to others and practical.

Many challenges still remain. 
Program synthesis is a diverse field ranging from machine learning to programming language communities. 
While we have modular blocks for various approaches, we continuously add new ones to express more ideas.



\acks{We would like to thank Neil Yorke-Smith and Dekel Zak for their valuable feedback and thoughtful review of our paper.
This project is funded by the Dutch Research Council (OCENW.M.21.325) and the National Growth Fund of the Ministry of Economic Affairs and the Ministry of Finance (Future Network Services).
}


\newpage




\vskip 0.2in
\bibliography{sample}

\end{document}